\documentstyle[eqsecnum,pre,aps,multicol,epsf]{revtex}
\begin{document}
\bibliographystyle{prsty}
\title{Dynamics of Viscoplastic Deformation in Amorphous 
Solids}
\author{M.L. Falk and J.S. Langer}
\address{Department of Physics, University of California,
        Santa Barbara, CA 93106}
\date{\today}
\maketitle

\begin{abstract}
We propose a dynamical theory of low-temperature shear deformation in
amorphous solids.  Our analysis is based on molecular-dynamics
simulations of a two-dimensional, two-component noncrystalline system. 
These numerical simulations reveal behavior typical of metallic glasses
and other viscoplastic materials, specifically, reversible elastic 
deformation at small applied stresses, irreversible plastic deformation 
at larger
stresses, a stress threshold above which unbounded plastic flow occurs,
and a strong dependence of the state of the system on the history of
past deformations.  Microscopic observations suggest that a dynamically
complete description of the macroscopic state of this deforming body
requires specifying, in addition to stress and strain, certain average
features of a population of two-state shear transformation zones.  Our
introduction of these new state variables into the constitutive
equations for this system is an extension of earlier models of creep in
metallic glasses.  In the treatment presented here, we specialize to
temperatures far below the glass transition, and postulate that
irreversible motions are governed by local entropic fluctuations in the
volumes of the transformation zones.  In most respects, our theory is
in good quantitative agreement with the rich variety of phenomena seen
in the simulations. \end{abstract}

\begin{multicols}{2}
\section{Introduction}
\label{sec:intro}

This paper is a preliminary report on a molecular-dynamics investigation 
of viscoplastic deformation in a non-crystalline solid.  It is 
preliminary in the sense that we have completed only the initial stages 
of our planned 
simulation project.  The results, however, have led us to a theoretical 
interpretation that we believe is potentially useful as a guide for 
further investigations along these lines.  In what follows, we describe 
both the 
simulations and the theory.  

Our original motivation for this project was an interest in the physics 
of deformations near the tips of rapidly advancing cracks, where 
materials are subject to very large stresses and experience very high 
strain rates.  Understanding the dissipative dynamics which occur in the 
vicinity of the 
crack tip is necessary to construct a satisfactory theory of dynamic 
fracture.\cite{JLanger97}  Indeed, we believe that the problem of dynamic fracture 
cannot be separated from the problem of understanding the 
conditions under which a solid behaves in a brittle or ductile manner.\cite{Rice74,Rice92,Khantha94,Freund90,Steif83}  To undertake 
such a project we eventually shall need sharper definitions of the terms 
``brittle'' and ``ductile'' than are presently available; but we leave 
such questions to future investigations
while we focus on the specifics of deformation in the absence of a 
crack.

We have chosen to study amorphous materials because the best experiments 
on dynamic instabilities in fracture have been carried out in silica 
glasses 
and polymers.\cite{Fineberg91,Fineberg92}   We know that amorphous materials exhibit both brittle and 
ductile behavior, often in ways that, on a macroscopic level, look very 
similar to deformation in crystals.\cite{Wu90}  More generally, we are looking for 
fundamental principles that might point us toward theories of 
deformation and failure in broad classes of macroscopically isotropic 
solids where thinking of deformation in terms of the dynamics of 
individual dislocations\cite{Rice74,Rice92} is either suspect, due to the absence of 
underlying crystalline order, or simply intractable, due to the extreme 
complexity of such an undertaking. In this way we hope that the ideas 
presented here will be generalizable perhaps to some 
polycrystalline materials or even single crystals with large numbers of 
randomly distributed dislocations.

We describe our numerical experiments in Section 
\ref{sec:computational}. Our working material is a two-dimensional,
two-component, noncrystalline solid in which the molecules interact via
Lennard-Jones forces.  We purposely maintain our system at a
temperature very far below the glass transition.  In the experiments,
we subject this material to various sequences of pure shear stresses,
during which we measure the mechanical response.  The simulations
reveal a rich variety of behaviors typical of metallic glasses\cite{Chaudhari83,Spaepen83,Taub82a,Kimura83} and other
viscoplastic solids,\cite{Oleinik93} specifically: reversible elastic
deformation at small applied stresses, irreversible plastic deformation
at somewhat larger stresses, a stress threshold above which unbounded
plastic flow occurs, and a strong dependence of the state of the system
on the history of past deformations.  In addition, the
molecular-dynamics method permits us to see what each molecule is doing
at all times; thus, we can identify the places where irreversible
molecular rearrangements are occurring. 

Our microscopic observations suggest that a dynamically 
complete description of the macroscopic state of this 
deforming body requires specifying, in addition to stress 
and strain, certain average features of a population of what 
we shall call ``shear transformation zones.''  These zones 
are small regions, perhaps consisting of only five or ten 
molecules, in special configurations that are particularly 
susceptible to inelastic rearrangements in response to shear 
stresses.  We argue that the constitutive relations for a 
system of this kind must include equations of motion for the 
density and internal states of these zones; that is, we must 
add new time-dependent state variables to the dynamical 
description of this system.\cite{Hart70,Rice75}  Our picture of shear 
transformation zones is based on earlier versions of the 
same idea due to Argon, Spaepen and others who described
creep in metallic alloys in terms of activated transitions 
in intrinsically heterogeneous 
materials.\cite{Spaepen77,Argon79a,Argon79b,Spaepen81,Argon83,Khonik94}  
These theories, in turn, drew on previous 
free-volume formulations of the glass transition by Turnbull, Cohen and
others in relating the transition rates to local free-volume
fluctuations.\cite{Spaepen81,Cohen59,Turnbull61,Turnbull70}  None of
those theories, however, were meant to describe the low-temperature
behavior seen here, especially the different kinds of irreversible
deformations that occur below and above a stress threshold, and the
history dependence of the response of the system to applied loads.

We present theory of the dynamics of shear transformation zones in Section 
\ref{sec:theory}.  This theory contains
four crucial features that are not, so far as we know, in any previous 
analysis:  First, once a zone has transformed and relieved a certain 
amount of shear stress, it cannot transform again in the same direction.  
Thus, the system saturates and, in the language of granular materials, 
it becomes ``jammed.''   Second, zones can be created and destroyed at 
rates proportional to the rate of irreversible plastic work being done 
on the system.  This is the ingredient that produces a threshold for 
plastic flow; the system can become ``unjammed'' when new zones are 
being created as fast as existing zones are being transformed.  Third, 
the attempt frequency is tied to the noise in the system, which is 
driven by the strain rate.  The stochastic nature of these fluctuations 
is assumed to arise from random motions associated with the disorder in 
the system.  And, fourth, the transition rates are strongly sensitive to 
the applied stress.  It is this sensitivity that produces memory 
effects.  

The resulting theory accounts for many of the features of the 
deformation dynamics seen in our simulations.  However, it is a mean 
field theory which fails to take into account any spatial correlations 
induced by interactions
between zones, and therefore it cannot explain all aspects of the 
behavior that we observe.  In particular, the mean-field nature of our 
theory precludes, at least for the moment, any analysis of strain 
localization or shear banding. 

\section{Molecular-Dynamics Experiments}
\label{sec:computational}

\subsection{Algorithm}
Our numerical simulations have been performed in the spirit of 
previous investigations of deformation in amorphous 
solids\cite{Deng89,Kobayashi80,Maeda81,Srolovitz83}.  We 
have examined the response to an applied shear of a noncrystalline, 
two-dimensional, two-component solid composed 
of either 10,000 or 20,000 molecules interacting via 
Lennard-Jones forces.  Our molecular dynamics (MD) algorithm 
is derived from a standard ``NPT'' dynamics 
scheme\cite{Melchionna93}, i.e. a pressure-temperature 
ensemble, with a Nose-Hoover thermostat,\cite{Nose84,Nose84b,Nose86}
and a Parinello-Rahman barostat\cite{Parrinello81,Parrinello82}
modified to allow imposition of an arbitrary two-dimensional stress
tensor.  The system obeys periodic boundary conditions, and both the
thermostat and barostat act uniformly throughout the sample.   

Our equations of motion are the following:
\begin{eqnarray}
\label{MDeq1}
{\bf \dot{r}}_n & = & \frac{{\bf p}_n}{m_n} +
[\dot{\varepsilon}]\cdot({\bf r}_n - {\bf R}_0)\\ 
\label{MDeq2}
{\bf \dot{p}}_n & = &
{\bf F}_n - ([\dot{\varepsilon}] + \xi [I]){\bf p}_n\\ 
\label{MDeq3}
\dot{\xi}       & =
& \frac{1}{\tau_T^2} (\frac{T_{kin}}{T} - 1) \\
\label{MDeq4}
\left[\ddot{\varepsilon}\right]    & = &
-\frac{1}{\tau_P^2}\frac{V}{Nk_BT} ([\sigma_{av}]-
[\sigma]) \\
\label{MDeq5}
{\bf \dot{L}} & = & [\dot{\varepsilon}]\cdot{\bf L}
\end{eqnarray}
Here, ${\bf r}_n$ and ${\bf p}_n$ are the position and 
momentum of the $n$'th molecule, and ${\bf F}_n$ is the force 
exerted on that molecule by its neighbors via the Lennard-
Jones interactions.  The quantities in brackets, e.g. 
$[\dot\varepsilon]$ or $[\sigma]$, are two-dimensional 
tensors.  $T$ is the temperature of the thermal reservoir; 
$V$ is the volume of the system (in this case, the area), 
and $N$ is the number 
of molecules.  $T_{kin}$ is the average kinetic energy per 
molecule divided by Boltzmann's constant $k_B$.  $[\sigma]$ 
is the externally applied stress, and $[\sigma_{av}]$ is the 
average stress throughout the system computed to be
\begin{equation}
[\sigma_{av}]_{ij} = {1 \over 4 V} \sum_{n} \sum_{m} 
F_{nm}^{i} r_{nm}^{j},
\end{equation} 
where $F_{nm}^i$ is the $i$'th component of the force between
particles $n$ and $m$; $r_{nm}^j$ is the $j$'th component of
the vector displacement between those particles; and $V$
is the volume of the system.  {\bf L} is the locus of points which
describe the boundary of the simulation cell.  While (\ref{MDeq5}) 
is not directly relevant to the dynamics of the particles, keeping
track of the boundary is necessary in order to properly calculate
intermolecular distances in the periodic cell.

The additional dynamical degrees of freedom in 
(\ref{MDeq1}-\ref{MDeq5}) are a viscosity $\xi$, which couples the
system to the thermal reservoir, and a strain rate,
$[\dot{\varepsilon}]$ via which the externally applied stress is
transmitted to the system.  Note that $[\dot{\varepsilon}]$ induces an
affine transformation about a reference point ${\bf R}_0$ which,
without loss of generality, we choose to be the origin of our
coordinate system.  In a conventional formulation, $[\sigma]$ would be
equal to $-P\,[I]$, where $P$ is the pressure and $[I]$ is the unit
tensor.  In that case, these equations of motion are known to produce
the same time-averaged equations of state as an equilibrium NPT
ensemble.\cite{Melchionna93}  By instead controlling the tensor $[\sigma]$, including its
off-diagonal terms, it is possible to apply a shear stress to the
system without creating any preferred surfaces which might enhance
system-size effects and interfere with observations of bulk properties.
The applied stress and the strain-rate tensor are constrained to be
symmetric in order to avoid physically uninteresting rotations of the
cell. Except where otherwise noted, all of our numerical experiments
are carried out at constant temperature, with $P\,=\,0$, and with the
sample loaded in uniform, pure shear.

We have chosen the artificial time constants $\tau_T$ and 
$\tau_P$ to represent physical aspects of the system.  As 
suggested by Nose\cite{Nose84}, $\tau_T$ is the time for a 
sound wave to travel an interatomic distance and, as 
suggested by Anderson\cite{Anderson80}, $\tau_P$ is the time 
for sound to travel the size of the system.

\subsection{Model Solid}

The special two-component system that we have chosen to 
study here has been the subject of other 
investigations\cite{Lancon86,Lancon88,Mikulla95} primarily 
because it has a quasi-crystalline ground state.  The 
important point for our purposes, however, is that this 
system can be quenched easily into an apparently stable 
glassy state.  Whether this is actually a thermodynamically 
stable glass phase is of no special interest here.  We care 
only that the noncrystalline state has a lifetime that is 
very much longer than the duration of our experiments.

Our system consists of molecules of two different sizes, 
which we call ``small'' ($S$) and ``large'' ($L$).  The 
interactions between these molecules are standard 6-12 
Lennard-Jones potentials:
\begin{equation}
\label{LJ}
U_{\alpha \beta}(r) = 4e_{\alpha
\beta}\left[\left(\frac{a_{\alpha \beta}}{r} \right)^{12}
-\left(\frac{a_{\alpha \beta}}{r}\right)^{6}\right] 
\end{equation}
where the subscripts $\alpha$, $\beta$ denote $S$ or $L$. We 
choose the zero-energy interatomic distances, $a_{\alpha 
\beta}$, to be
\begin{equation} 
a_{SS} = 2 \sin(\frac{\pi}{10}),~~~~ a_{LL} = 
2\sin(\frac{\pi}{5}),~~~~ a_{SL} = 1;
\end{equation} 
with bond strengths:
\begin{equation}
e_{SL} = 1,~~~~ e_{SS} = e_{LL} =\frac{1}{2}.
\end{equation}
For computational efficiency, we impose a finite-range 
cutoff on the potentials in (\ref{LJ}) by setting them equal 
to zero for separation distances $r$ greater than $2.5 
a_{SL}$.  The masses are all taken to be equal. The ratio of 
the number of large molecules to the number of small 
molecules is half the golden mean: 
\begin{equation}
\frac{N_L}{N_S} =\frac{1+\sqrt{5}}{4}. 
\end{equation} 
In the resulting system, it is energetically favorable for 
ten small molecules to surround one large molecule, 
or for five large molecules to surround one small molecule.  The 
highly frustrated nature of this system avoids problems of 
local crystallization that often occur in two dimensions 
where the nucleation of single component crystalline regions 
is difficult to avoid.  As shown by Lan\c{c}on et 
al\cite{Lancon86}, this system goes through something 
like a glass transition upon cooling from its liquid state.  
The glass transition temperature is $0.3 T_0$ where 
$k_B\,T_0 = e_{SL}$.  All the simulations reported here have 
been carried out at a temperature $T=0.001\,T_0$, that is, 
at 0.3\% of the glass transition temperature.  Thus, all of 
the phenomena to be discussed here take place at a 
temperature very much lower than the energies associated 
with the molecular interactions. 

In order to start with a densely packed material, we have 
created our experimental systems by equilibrating a random 
distribution of particles under high pressure at the low 
temperature mentioned above. After allowing the system to 
relax at high pressure, we have reduced the pressure to zero and 
again allowed the sample to relax. Our molecular dynamics 
procedure permits us to relax the system only for times of 
order nanoseconds, which are not long enough for the 
material to experience any significant amount of annealing, 
especially at such a low temperature.

We have performed numerical experiments on two different 
samples, containing 10,000 and 20,000 molecules 
respectively.  All of the simulation results
shown are from the larger of the two samples; the smaller sample has
been used primarily to check the reliablility of our procedures.
We have created each of these samples only 
once; thus each experiment using either of them starts with 
precisely the same set of molecules in precisely the same 
positions.  As will become clear, there are both advantages 
and uncertainties associated with this procedure.  On the 
one hand, we have a very carefully controlled starting point 
for each experiment.  On the other hand, we do not know how 
sensitive the mechanical properties of our system 
\end{multicols}
\nopagebreak
\begin{center}
\begin{table}[]
\begin{minipage}[t]{5in}
\begin{tabular}{|c||c|c|c|c|c|}
   & Molecules & Shear Modulus & Bulk Modulus & 2D Poisson Ratio &
   Young's Modulus\\ \hline
 Sample 1 & 10,000    & 9.9 & 31 & 0.51 & 30\\ \hline
 Sample 2 & 20,000    & 16  & 58 & 0.57 & 50
\label{tab1}
 \end{tabular}
\caption{Sample Sizes and Elastic Constants}
\end{minipage}
\end{table}
\end{center}
\begin{multicols}{2}
\noindent might be 
to details of the preparation process, nor do we know 
whether to expect significant sample-to-sample variations in 
the molecular configurations.  To illustrate these 
uncertainties, we show the elastic constants of the samples 
in Table 1. The moduli are expressed there in units of 
$e_{SL}/a_{SL}^2$.  (Note that the Poisson ratio for a two-
dimensional system has an upper bound of 1 rather than 0.5 
as in the three-dimensional case.)  The appreciable 
differences between the moduli of supposedly identical 
materials tell us that we must be very careful in drawing 
detailed conclusions from these preliminary results.

\subsection{Simulation Results: Macroscopic Observations}

In all of our numerical experiments, we have tried simply to mimic 
conventional laboratory measurements of viscoplastic properties of real 
materials.  The first of these is a measurement of stress at constant 
strain rate.  As we shall see, this supposedly simplest of the 
experiments is  especially interesting and problematic for us because it 
necessarily probes time-dependent behavior near the plastic yield 
stress.  

\begin{figure}[t]
\epsfxsize=3.0in
\centerline{\epsfbox{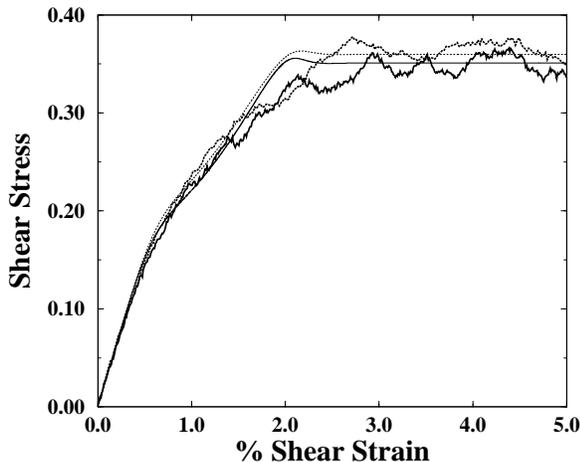}}
\begin{minipage}[t]{8.1cm} 
\caption{ \label{strainrate} 
Shear stress vs. strain for strainrates of $10^{-4}$ (solid lines) and 
$2 \times 10^{-4}$ (dotted lines).  The thicker lines which denote the
simulation results exhibit both linear elastic behavior at low
strain and non-linear response leading to yield at approximately
$\sigma_s = 0.35$.  The thinner curves are predictions of the theory for
the two strain rates.  Strainrate is measured in units of 
$(e_{SL} / m\,a^2_{SL})^{1 \over 2}$.}
\end{minipage} 
\end{figure}

Our results, for two different strain rates, are shown in 
Figure~\ref{strainrate}.  The strain rates are expressed in units proportional 
to the frequency of oscillation about the minimum in the Lennard-Jones 
potential, specifically, in units of 
$\omega_0\equiv (e_{SL} /m a_{SL}^2 )^{1\over2}$, where $m$ is the 
particle mass. (The actual frequency for the SL potential, in cycles per 
second, is $(3\cdot 2^{1/3} / \pi)\,\omega_0\cong 1.2\,\omega_0$.) As 
usual, the sample has been kept at constant temperature and at pressure 
$P=0$. At low strain, the material behaves in a linearly elastic manner.  
As the strain increases, the response   
\begin{figure}[t]
\epsfxsize=3.0in
\centerline{\epsfbox{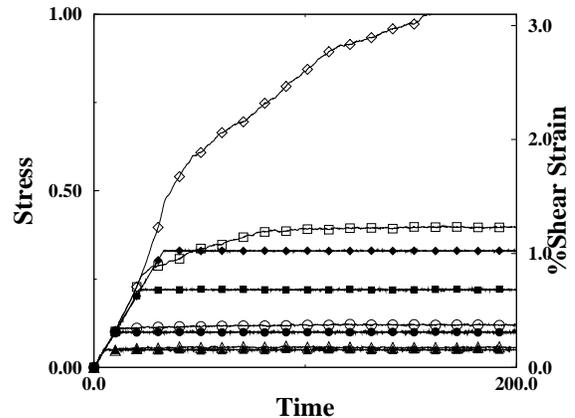}}
\begin{minipage}[t]{8.1cm} 
\caption{ \label{multistrain}
Shear strain (open symbols) vs. time for
several applied shear stresses (solid symbols).  The stresses have been
ramped up at a constant rate until reaching a maximum value and then have been
held constant. The strain and stress axes are related by twice the
shear modulus so that, for linear elastic response, the open and
closed symbols would be coincident. For low stresses the sample
responds in an almost entirely elastic manner.  For intermediate
stresses the sample undergoes some plastic deformation prior to
jamming.  In the case where the stress is brought above the
yield stress, the sample deforms indefinitely. Time is measured in units
of $( m a_{SL}^2 / e_{SL})^{1 \over 2}$.}
\end{minipage} 
\end{figure}
\noindent becomes nonlinear, and the 
material begins to deform plastically.  Plastic yielding, that is, the 
onset of plastic flow, occurs when the strain reaches approximately 0.7\%.  
Note that the stress does not rise smoothly 
and monotonically in these experiments.  We presume that most of this 
irregularity would average out in larger systems.  As we shall see, 
however, there may also be more interesting dynamical effects at work 
here.

In all of the other experiments to be reported here, we have 
controlled the stress on the sample and measured the strain.  
In the first of these, shown in Figure~\ref{multistrain}, we have
increased the stress to various different values and then held it
constant. 

In each of these experimental runs, the stress starts at zero
and increases at the same constant rate until the desired final stress
is reached.  The graphs show both this applied stress (solid symbols)
and the resulting strain (open symbols), as functions of time, for
three different cases.  Time is measured in the same
molecular-vibration units used in the previous experiments, i.e. in
units of $(m a_{SL}^2 / e_{SL})^{1 \over 2}$.  The stresses and strain
axes are related by twice the shear modulus so that, if the response is
linearly elastic, the two curves lie on on top of one another. In the
case labelled ($\triangle$), the final stress is small, and the
response is nearly elastic.  For cases ($\circ$) and ($\Box$), the
sample deforms plastically until it reaches some final strain, at which
it ceases to undergo further deformation 
\begin{figure}
\epsfxsize=3.0in
\centerline{\epsfbox{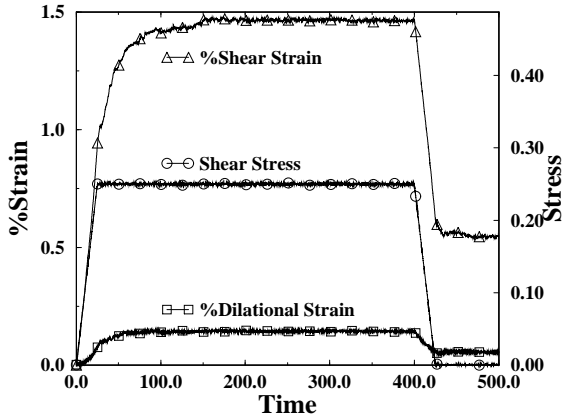}}
\begin{minipage}[t]{8.1cm} 
\caption{ Stress and strain vs. time for one
particular loading where the stress has been ramped up to $\sigma_s = 0.25$,
held for a time, and then released.  Note that, in addition to the shear 
response, the material undergoes a small amount of dilation. 
}\label{strain-vs-time} 
\end{minipage}
\end{figure}
\begin{figure}
\epsfxsize=3.0in
\centerline{\epsfbox{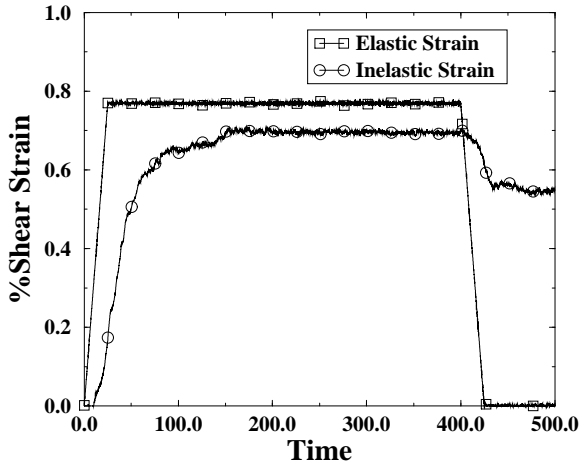}}
\begin{minipage}[t]{8.1cm} 
\caption{\label{inelastic}
Elastic and inelastic strain vs time for the same simulation as 
that shown in Figure~\ref{strain-vs-time}.  The inelastic strain is 
found by subtracting the linearly elastic strain from the total strain.  
Note the partial  recovery of the inelastic portion of the strain which 
occurs during and after unloading.
}
\end{minipage}
\end{figure}
\noindent on observable timescales.  (We
cannot rule out the possibility of slow creep at much longer times.) 
In case ($\Diamond$), for which the final stress is the largest of the
three cases shown, the sample continues to deform plastically at
constant stress throughout the duration of the experiment.  
We conclude from these and a number of similar experimental 
runs that there exists a well defined critical stress for 
this material, below which it reaches a limit of plastic 
deformation, that is, it ``jams,'' and above which it flows 
plastically.  
Because the stress is ramped up quickly, we can see in 
curves ($\Box$) and ($\Diamond$) of Figure~\ref{multistrain} that there
is a separation of time scales between the elastic  and plastic
responses.  The elastic response is instantaneous, while the plastic
response develops over a few hundred molecular vibrational periods.
To see the distinction between these behaviors more clearly, we have
performed experiments in which we load the system to a fixed,
subcritical stress, hold it there, and then unload it by ramping the
stress back down to zero. In Figure~\ref{strain-vs-time}, we show this
stress and the resulting total shear strain, as functions of time, for
one of those experiments.  If we define the the
elastic strain to be the stress divided by twice the previously
measured, as-quenched, shear modulus, then we can compute the inelastic
strain by subtracting the elastic from the total.  The result is shown
in Figure~\ref{inelastic}.  Note that most, but not quite all, of the
inelastic strain consists of nonrecoverable plastic deformation that
persists after unloading to zero stress.  Note also, as shown in
Figure~\ref{strain-vs-time}, that the system undergoes a small dilation
during this process, and that this dilation appears to have both
elastic and inelastic components.

\begin{figure}
\epsfxsize=2.5in
\centerline{\epsfbox{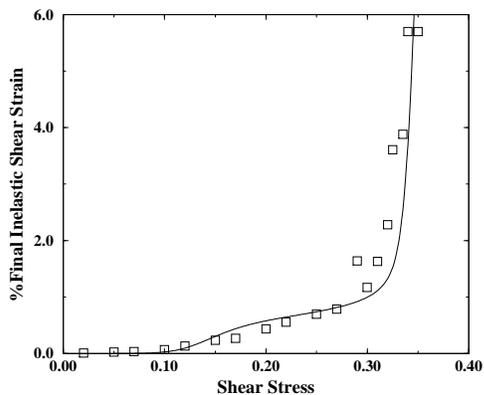}}
\begin{minipage}[t]{8.1cm} 
\caption{Final inelastic strain vs. applied stress for
stresses below yield.  The simulation data (squares) have been obtained by 
running the simulations until all deformation apparently had stopped. 
The comparison to the theory (line) was obtained by numerically 
integrating the equations of motion for a period of 800 time units, the 
duration of the longest simulation runs.}\label{finalstrain}
\end{minipage}
\end{figure}

Using the simple prescription outlined above, we have 
measured the final  inelastic shear strain as a function of 
shear stress.  That is, we have measured the shear strain 
once the system has ceased to deform as in the subcritical 
cases in Figure~\ref{multistrain}, and then subtracted the elastic
part.  The results are shown in Fig.~\ref{finalstrain}.  As expected,
we see only very small amounts of inelastic strain at low stress.  As
the stress approaches the yield stress, the inelastic strain appears to
diverge approximately logarithmically.

\begin{figure}
\epsfxsize=3.0in
\centerline{\epsfbox{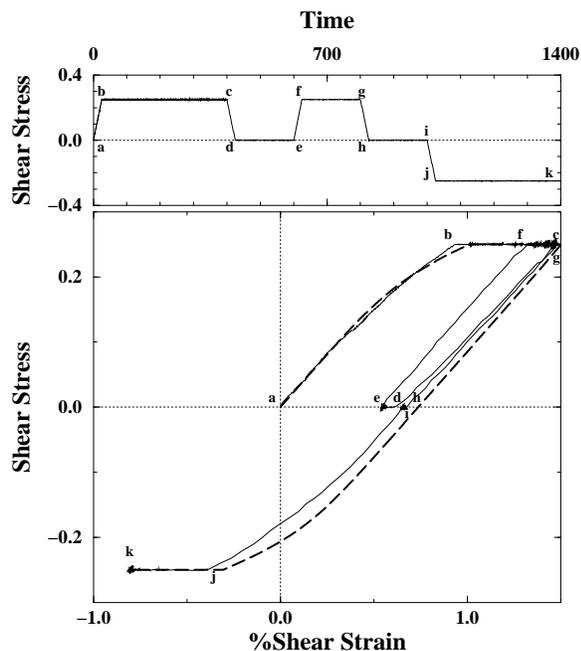}}
\begin{minipage}[t]{8.1cm} 
\caption{Stress-strain trajectory for a molecular\-dynamics experiment in 
which the sample has been loaded, unloaded, reloaded, unloaded again,
and then reverse-loaded, all at stresses below the yield stress.  The 
smaller graph above shows the history of applied shear stress with 
letters indicating  identical times in the two graphs.  The dashed line 
in the main graph is the  theoretical prediction for the same sequence 
of stresses. Note that a small amount of inelastic strain recovery 
occurs after the first unloading in the simulation, but that no such 
behavior occurs in the theory.  Thus, the theoretical curve from {\bf c} 
though {\bf h} unloads, reloads and unloads again all along the same 
line.}\label{cycle}
\end{minipage}
\end{figure}

The final test that we have performed is to cycle the system 
through loading, reloading, and reverse-loading. As shown in 
Figure~\ref{cycle}, the sample is first loaded on the curve 
from {\bf a} to {\bf b}.  The initial response is linearly 
elastic, but, eventually, deviation from linearity occurs as 
the material begins to deform inelastically. From {\bf b} to 
{\bf c}, the stress is constant and the sample continues to 
deform inelastically until reaching a final strain at {\bf 
c}. Upon unloading, from {\bf c} to {\bf d}, the system does 
not behave in a  purely elastic manner but, rather, recovers some
portion of the strain anelastically.  While held at 
zero stress, the sample continues to undergo anelastic 
strain recovery from {\bf d} to {\bf e}.

When the sample is then reloaded from {\bf e} to {\bf f}, it 
undergoes much less inelastic deformation than during the 
initial loading.  From {\bf f} to {\bf g} the sample again 
deforms inelastically, but by an amount only slightly more 
than the previously recovered strain, returning 
approximately to point {\bf c}.  Upon unloading again from 
{\bf g} to {\bf h} to {\bf i}, less strain is recovered than 
in the previous unloading from {\bf c} through {\bf e}. 

It is during reverse loading from {\bf i} to {\bf k} that it 
becomes
apparent that the deformation history has rendered the 
amorphous sample
highly anisotropic in its response to further applied shear. 
The inelastic strain from {\bf i} to {\bf k} is much greater 
than that from {\bf e} to {\bf g}, demonstrating a very 
significant Bauschinger effect.  The plastic deformation in 
the initial direction apparently has biased the sample in 
such a way as to inhibit further inelastic yield 
\noindent in the same 
direction, but there is no such inhibition in the reverse 
direction. The material, therefore, must in some way have 
microstructurally encoded, i.e. partially ``memorized,'' its 
loading history.

\subsection{Simulation Results: Microscopic Observations}

Our numerical methods allow us to examine what is happening at the 
molecular level during these deformations.  To do this systematically, 
we need to identify where irreversible plastic rearrangements are 
occurring.  More precisely, we must identify places where the molecular 
displacements are non-affine, that is, where they deviate substantially 
from displacements which can be described by a linear strain field.  
Our mathematical technique for identifying regions of non-affine 
displacement has been described by one of us (MLF) in an earlier 
publication.\cite{Falk98}  For completeness, we repeat it here.

We start with a set of molecular positions and subsequent displacements, 
and compute the closest possible approximation to a local strain tensor 
in the neighborhood of any particular molecule.  To define that 
neighborhood, we  define a sampling radius, which we choose to be the 
interaction range, $2.5 a_{SL}$.  The local strain is then determined by 
minimizing the mean 
square difference between the the actual displacements of the 
neighboring molecules relative to the central one, and the relative 
displacements that they would have if they were in a region of uniform 
strain $\varepsilon_{ij}$.  That is, we define
\begin{eqnarray} 
\lefteqn{ D^2(t, \Delta t) =  \sum_{n}\sum_{i}\Bigl[r^{i}_n(t)-r^{i}_0(t) - \Bigr. }\nonumber\\ 
& & \Bigl. \sum_{j} (\delta_{ij} + \varepsilon_{ij}) \Bigl(r^{j}_n(t-\Delta t) - 
r^{j}_0(t-\Delta t)\Bigr)
 \Bigr]^2,
\end{eqnarray}
where the indices $i$ and $j$ denote spatial coordinates, and the index 
$n$ runs over the molecules within the interaction range of the 
reference molecule, $n=0$ being the reference molecule. $r^i_n(t)$ is 
the $i$'th component of the position of the $n$'th molecule at time $t$.  
We then find the $\varepsilon_{ij}$ which minimizes $D^2$ by 
calculating: 
\begin{eqnarray} 
X_{ij} &=
\sum_{n} &(r^{i}_n(t)-r^{i}_0(t)) \times \nonumber \\
& & (r^{j}_n(t-\Delta t)-r^{j}_0(t-\Delta t)), \\ \nonumber\\
 Y_{ij} & = 
\sum_{n}&(r^{i}_n(t-\Delta t)-r^{i}_0(t-\Delta t)) \times \nonumber\\
 & & (r^{j}_n(t-\Delta t)-r^{j}_0(t-\Delta t)), \\ \nonumber\\
\varepsilon_{ij} & =  \sum_{k}& X_{ik} Y_{jk}^{-1} -
\delta_{ij}. 
\end{eqnarray} 
The minimum value of $D^2(t, \Delta t)$ is then the local deviation from 
affine deformation during the time interval $[t-\Delta t,\,t]$. We shall 
refer to this quantity as $D_{min}^2$.

\begin{figure}
\epsfxsize=3.0in
\centerline{\epsfbox{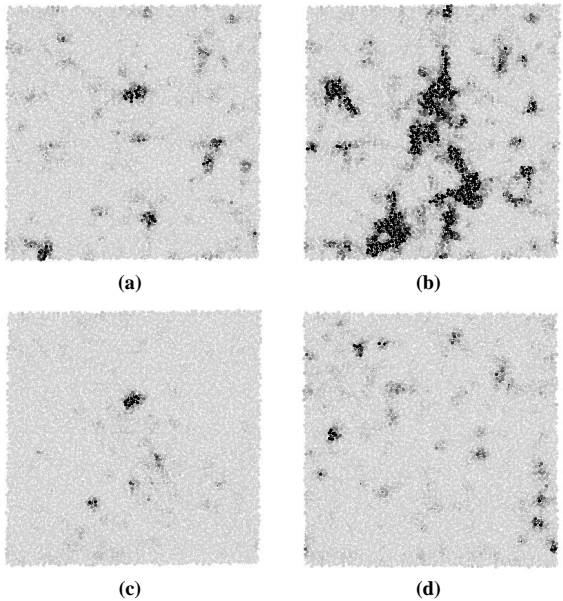}}
\begin{minipage}[t]{8.1cm} 
\caption{Intensity plots of $D_{min}^2$, the deviation from affine deformation,
for various intervals during two
simulations.  Figures (a), (b) and (c) show deformation during one simulation
in which the stress has been ramped up quickly to a value less than the yield 
stress and then held constant.  Figure (a) shows deformations over the first 10
time units, and figure (b) over the first 30 time units.  Figure (c) shows
the same state as in (b), but with $D_{min}^2$ computed only for deformations
that took place during the preceeding 1 time unit.  In Figure (d), the 
initial system and the time interval (10 units) are the same as in (a),
but the stress has been applied in the opposite direction.  The gray scale in
these figures has been selected so that the darkest spots identify molecules
for which $|D_{min}| \approx 0.5\,a_{SL}$.}\label{STzones1}
\end{minipage}
\end{figure}
\begin{figure}
\epsfxsize=3.0in
\centerline{\epsfbox{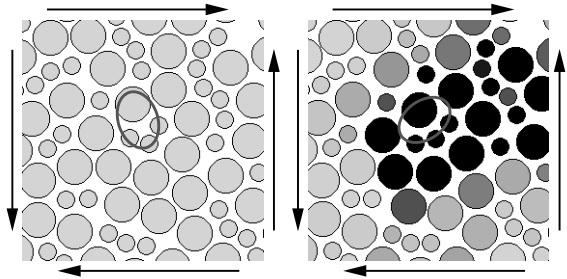}}
\begin{minipage}[t]{8.1cm} 
\caption{Close-up picture of a shear transformation zone before and after
undergoing transformation.  Molecules after transformation are shaded according
to their values of $D_{min}^2$ using the same gray scale as in 
Figure \ref{STzones1}.  The direction of the externally applied
shear stress is shown by the arrows.  The ovals are included solely as guides
for the eye.}\label{STzones2}
\end{minipage}
\end{figure}
\vspace{0.5in}

We have found that $D_{min}^2$ is an excellent diagnostic for 
identifying local irreversible shear transformations.  Figure 
\ref{STzones1} contains four different intensity plots of $D_{min}^2$ 
for a particular system as it is undergoing plastic deformation.  The 
stress has been ramped up to $|\sigma_s| = 0.12$ in the time interval 
[0,12] and then held constant in an experiment analogous to that shown in 
Figure \ref{multistrain}.  Figure \ref{STzones1}(a) shows $D_{min}^2$ for 
$t=10$, $\Delta t =10$. It demonstrates that the non-affine deformations 
occur as isolated small events.  In (b) we observe the same simulation,
but for $t=30$, $\Delta t =30$; that is, we are looking at a later time, but 
again we consider rearrangements relative to the inital configuration.  
Now it appears that the regions of rearrangement
have a larger scale structure.  The pattern seen here looks like an 
incipient shear band.  However, in (c), where $t=30$, $\Delta t = 1$,
we again consider this later time but look only at
rearrangements that have occurred in the preceeding short
time interval.  The events shown in this figure are small, demonstrating
that the pattern shown in (b) is, in fact, an aggregation of many local
events.  Lastly, in (d), we show an experiment similar in all respects to
(a) except that the sign of the stress has been reversed.  As in
(a), $t=10$, $\Delta t =10$, and again we observe small isolated events. 
However, these events occur in different locations, implying a 
direction dependence of the local transformation mechanism.

Next we look at these processes in yet more detail. Figure 
\ref{STzones2} is a close-up of the molecular configurations in 
the lower left-hand part of the largest dark cluster seen in 
Figure \ref{STzones1}(c), shown just before and just after a shear 
transformation.  During this event, the cluster of one large and three
small molecules has compressed along the top-left/bottom-right axis 
and extended along the bottom-left/top-right axis.  This deformation is 
consistent with the orientation of the applied shear, which is in the 
direction shown by the arrows on the outside of the figure. Note that 
this rearrangement takes place without significantly affecting the 
relative positions of molecules in the immediate environment of the 
transforming region.  This is the type of 
rearrangement that Spaepen identifies as a 
``flow defect.'' \cite{Spaepen81}  As mentioned in the introduction, we 
shall call these regions ``shear transformation zones.''

\section{Theoretical Interpretation of the Molecular-Dynamics 
Experiments}
\label{sec:theory}

\subsection{Basic Hypotheses}

We turn now to our attempts to develop a theoretical 
interpretation of the phenomena seen in the simulations.  We 
shall not insist that our theory reproduce every detail of 
these results.  In fact, the simulations are not yet 
complete enough to tell us whether some of our observations 
are truly general properties of the model or are artifacts 
of the ways in which we have prepared the system and carried 
out the numerical experiments.  Our strategy will be, first, 
to specify what we believe to be the basic framework of a 
theory, and then to determine which specific assumptions 
within this framework are consistent with the numerical 
experiments.

There are several features of our numerical experiments that 
we shall assume are fundamentally correct and which, 
therefore, must be outcomes of our theory.  These are: (1) 
At a sufficiently small, fixed load, i.e. under a constant 
shear stress less than some value that we identify as a 
yield stress, the system undergoes a finite plastic 
deformation.  The amount of this deformation diverges as the loading 
stress approaches the yield stress.  (2) At loading stresses above the 
yield 
stress, the system flows visco-plastically.  (3) The 
response of the system to loading is history-dependent.  If 
it is loaded, unloaded, and then reloaded to the same 
stress, it behaves almost elastically during the reloading, 
i.e. it does not undergo additional plastic deformation.  On 
the other hand, if it is loaded, unloaded, and then reloaded 
with a stress of the opposite sign, it deforms substantially 
in the opposite direction.  

Our theory consists of a set of rate equations describing plastic 
deformation.  These include an equation for the inelastic strain rate as 
a function of the stress plus other variables that describe the internal 
state of the system.  We also postulate equations of motion for these 
state variables.  Deformation theories of this type are in the spirit of 
investigations by E. Hart\cite{Hart70} who, to the best of our 
knowledge, was the first to argue in a mathematically systematic way 
that any satisfactory theory of plasticity must include dynamical state 
variables, beyond just stress and strain.  A similar point of view has
been stressed by Rice.\cite{Rice75}  Our analysis is also 
influenced by the use of state variables in theories of friction 
proposed recently by Ruina, Dieterich, Carlson, and others.\cite{Dieterich78,Dieterich79,Rice83,Ruina83,Dieterich94,Carlson96}

Our picture of what is happening at the molecular level in these systems 
is an
extension of the ideas of Turnbull, Cohen, Argon, Spaepen and 
others.\cite{Spaepen77,Argon79a,Argon79b,Spaepen81,Argon83,Cohen59,Turnbull61,Turnbull70}
These authors postulated that 
deformation in amorphous materials occurs at special sites where the 
molecules are able to rearrange themselves in response to applied 
stresses.  As described in the preceding chapter, we do see such sites 
in our simulations, and shall use these shear transformation zones 
as the basis for our analysis.  However, we must be careful to state as 
precisely as possible our definition of these zones, because we shall 
use them in ways that were not considered by the previous authors.

One of the most fundamental differences between previous 
work and ours is the fact that our system is effectively at 
zero temperature.  When it is in mechanical equilibrium, no 
changes occur in its internal state because there is no 
thermal noise to drive such changes.  Thus the shear 
transformation zones can undergo transitions only when the 
system is in motion.  Because the system is strongly 
disordered, the forces induced by large-scale motions at the 
position of any individual molecule may be noisy.  These 
fluctuating forces may even look as if they have a thermal 
component.\cite{SLanger97} The thermodynamic analogy 
(thermal activation of shear transformations with 
temperature being some function of the shear rate) may be an 
alternative to (or an equivalent of) the theory to be 
discussed here.  However, it is beyond the scope of the 
present investigation.

Our next hypothesis is that shear transformation zones are 
geometrically identifiable regions in an amorphous solid.  
That is, we assume that we could --- at least in principle
--- look at a picture of any one of the computer-generated 
states of our system and identify small regions that are 
particularly susceptible to inelastic rearrangement.  As 
suggested by Fig. \ref{STzones2}, these zones might consist of groups 
of four or more relatively loosely bound molecules surrounded by 
more rigid ``cages.''  But that specific picture is not necessary.  
The main idea is that some such irregularities are locked in 
on time scales that are very much longer than molecular 
collision times.  That is not to say that these zones are 
permanent features of the system on experimental time 
scales.  On the contrary, the tendency of these zones to 
appear and disappear during plastic deformation will be an 
essential ingredient of our theory. 

We suppose further that these shear transformation zones are 
two-state systems.  That is, in the absence of any 
deformation of the cage of molecules that surrounds them, 
they are equally stable in either of two configurations.  
Very roughly speaking, the molecular arrangements in these 
two configurations are elongated along one or the other of 
two perpendicular directions which, shortly, we shall take 
to be coincident with the principal axes of the applied 
shear stress. The transition between one such state and the 
other constitutes an elementary increment of shear strain.  
Note that bistability is the natural assumption here.  More 
than two states of comparable stability might be possible 
but would have relatively low probability.  A crucial 
feature of these bistable systems is that they can transform 
back and forth between their two states but cannot make 
repeated transformations in one direction.  Thus there is a 
natural limit to how much shear can take place at one of 
these zones so long as the zone remains intact.

We now consider an ensemble of shear transformation zones 
and estimate the probability that any one of them will 
undergo a transition at an applied shear stress $\sigma_s$.  
Because the temperatures at which we are working are so low 
that ordinary thermal activation is irrelevant, we focus our 
attention on entropic variations of the local free volume.  
Our basic assumption is that the transition probability is 
proportional to the probability that the molecules in a zone 
have a sufficiently large excess free volume, say $\Delta 
V^*$, in which to rearrange themselves.  This critical free 
volume must depend on the magnitude and orientation of the elastic 
deformation of the zone that is caused by the externally applied stress, 
$\sigma_s$. 

At this point, our analysis borrows in its general approach, but not in 
its specifics, from recent developments in the theory 
of granular materials\cite{Mehta90} where the 
only extensive state variable is the volume $\Omega$.  What follows is a 
very simple approximation which, at great loss of generality, leads us 
quickly to the result that we need.  The 
free volume, i.e. the volume in excess of close packing that 
the particles have available for motion, is roughly 
\begin{equation}
\Omega- N\,v_0\equiv N\,v_f,
\end{equation}
where $N$ is the total number of particles, $v_f$ is the 
average free volume per particle, and $v_0$ is the volume 
per particle in an ideal state of random dense packing.  In 
the dense solids of interest to us here, $v_f\ll v_0$, and 
therefore $v_0$ is approximately the average volume per 
particle even when the system is slightly dilated. The 
number of states available to this system is roughly 
proportional to $(v_f/h)^N$, where $h$ is an arbitrary 
constant with dimensions of volume --- the analog of 
Planck's constant in classical statistical mechanics --- 
that plays no role other than to provide dimensional 
consistency. Thus the entropy, defined here to be a 
dimensionless quantity, is
\begin{equation}
S(\Omega,N)\cong N\,\ln\Bigl({v_f\over h}\Bigr)\cong 
N\,\ln\left({\Omega - Nv_0\over N\,h}\right).
\end{equation}
The intensive variable analogous to temperature is $\chi$:
\begin{equation}
{1\over\chi}\equiv {\partial S\over 
\partial\Omega}\cong{1\over v_f}.
\end{equation}
Our activation factor, analogous to the Boltzmann factor for 
thermally activated processes, is therefore 
\begin{equation}
\label{expV}
e^{-(\Delta V^*/ \chi)}\cong e^{-(\Delta V^*/v_f)}.
\end{equation}

A formula like (\ref{expV}) appears in various places in the 
earlier literature.\cite{Spaepen77,Cohen59,Turnbull61,Turnbull70} 
There is an important difference between its earlier use and the way in
which we are using it here.  In earlier interpretations, (\ref{expV})
is an estimate of the probability that any given molecule has a large
enough free volume near it to be the site at which a thermally
activated irreversible transition might occur.  In our interpretation,
(\ref{expV}) plays more nearly the role of the thermal activation
factor itself.  It tells us something about the configurational
probability for a zone, not just for a single molecule.  When
multiplied by the density of zones and a rate factor, about which we
shall have more to say shortly, it becomes the transformation rate per
unit volume.  

Note what is happening here.  Our system is extremely non-ergodic
and, even when it is undergoing appreciable 
strain, does not explore more than a very small part of its 
configuration space.  Apart from the molecular 
rearrangements that take place during plastic deformation, 
the only chance that the system has for coming close to any 
state of equilibrium occurs during the quench by which it is 
formed initially.  Because we control only the temperature 
and pressure during that quench, we must use entropic 
considerations to compute the relative probabilities of 
various molecular configurations that result from it.  

The transitions occurring within shear transformation zones 
are strains, and therefore they must, in principle, be 
described by tensors.  For present purposes, however, we can 
make some simplifying assumptions.  As described in Chapter 
\ref{sec:computational}, our molecular-dynamics model is 
subject only to a uniform, pure shear stress of magnitude 
$\sigma_s$ and a hydrostatic pressure $P$ (usually zero).  
Therefore, in the principal-axis system of coordinates, the 
stress tensor is:
\begin{equation}
[\sigma] = \left[ \begin{array}{cc}
    -P 	   & \sigma_{s}\\ \sigma_{s} & -P
   \end{array}\right].
\end{equation}
Our assumption is that the shear transformation zones are 
all oriented along the same pair of principal axes, and 
therefore that the strain tensor has the form: 
\begin{equation} 
[\varepsilon] = \left[	\begin{array}{cc}
    \varepsilon_{d}&\varepsilon_{s}\\
    \varepsilon_{s}&\varepsilon_{d}
   \end{array}\right].
\end{equation}
where $\varepsilon_s$ and $\varepsilon_d$ are the shear and 
dilational strains respectively.  The total shear strain is 
the sum of elastic and inelastic components:
\begin{equation} 
\varepsilon_s = \varepsilon^{el}_s + \varepsilon^{in}_s.
\end{equation} 
By definition, the elastic component is the linear response 
to the stress:
\begin{equation}
\varepsilon^{el}_s = \frac{\sigma_s}{2 \mu},
\end{equation}
where $\mu$ is the shear modulus. 

In a more general 
formulation, we shall have to consider a distribution of 
orientations of the shear transformation zones.  That 
distribution will not necessarily be isotropic when plastic 
deformations are occurring, and very likely the distribution 
itself will be a dynamical entity with its own equations of 
motion.  Our present analysis, however, is too crude to 
justify any such level of sophistication.

The last of our main hypotheses is an equation of motion for 
the densities of the shear transformation zones.  Denote the 
two states of the shear transformation zones by the symbols 
$+$ and $-$, and let $n_{\pm}$ be the number densities of 
zones in those states.  We then write:
\begin{equation}
\label{ndot}
\dot n_{\pm}= R_{\mp}\,n_{\mp}-R_{\pm}\,n_{\pm} - 
C_1\,(\sigma_s\,\dot\varepsilon_s^{in})\,n_{\pm} 
+C_2\,(\sigma_s\,\dot\varepsilon_s^{in}).
\end{equation}
Here, the $R_{\pm}$ are the rates at which $\pm$ states transform to 
$\mp$ states.  These must be consistent with the transition 
probabilities described in the preceding paragraphs.  

The last two terms in (\ref{ndot}) describe the way in which the 
population of shear transformation zones changes as the system undergoes 
plastic deformation.  The zones can be annihilated and created --- as 
shown by the terms with coefficients $C_1$ and $C_2$ respectively --- at 
rates proportional to the rate $\sigma_s\dot\varepsilon_s^{in}$ at which 
irreversible work is being done on the system.  This last assumption is 
simple and plausible, but it is not strictly dictated by the physics in 
any way that we can see.  A caveat: In certain
circumstances, when the sample does work on its environment, 
$\sigma_s\dot\varepsilon_s^{in}$ could be negative, in which case the 
annihilation and creation terms in (\ref{ndot}) could produce results 
which would not be physically 
plausible.  We believe that such states in our theory are dynamically 
accessible only from unphysical starting configurations.  In related 
theories, however, that may not be the case.

It is important to recognize that the annihilation and creation terms in 
(\ref{ndot}) are interaction terms, and that they have been introduced 
here in a mean-field approximation.  That is, we implicitly assume that 
the rates at which shear transformation zones are annihilated and 
created depend only on the rate at which irreversible work is being done 
on the system as a whole, and that there is no correlation between the 
position at which the work is being done and the place where the 
annihilation or creation is occurring.  This is in fact not the case as
shown by Fig. \ref{STzones1}(b), and is possibly the weakest 
aspect of our theory.  

With the preceding definitions, the time rate of change of 
the inelastic shear strain, $\dot\varepsilon_s^{in}$, has 
the form:
\begin{equation}
\label{epdot1}
\dot\varepsilon_s^{in}= V_z\,\Delta\varepsilon\,\left[R_+\,n_
+-R_-\,n_-\right],
\end{equation}
where $V_z$ is the typical volume of a zone and 
$\Delta\varepsilon$ is the 
increment of local shear strain.

\subsection{Specific Assumptions}

We turn now to the more detailed assumptions and analyses 
that we need in order to develop our general hypotheses into 
a testable theory.

According to our hypothesis about the probabilities of 
volume fluctuations, we should write the transition rates in 
(\ref{ndot}) in the form:
\begin{equation}
\label{epdot}
R_{\pm}=R_0 \,\exp\left[-{\Delta V^*(\pm\sigma_s)\over v_f} 
\right].
\end{equation}
The prefactor $R_0$ is an as-yet unspecified attempt 
frequency for these transformations. In writing 
(\ref{epdot}), we have used the assumed symmetry of the 
system to note that, if $\Delta V^*(\sigma_s)$ is the 
required excess free volume for a $ (+ \to - )$ transition, 
then the appropriate free volume for the reverse transition 
must be $\Delta V^*(-\sigma_s)$.  We adopt the convention 
that a positive shear stress deforms a zone in such a way 
that it enhances the probability of a $(+ \to -)$ transition 
and decreases the probability of a $(-\to +)$ transition.  
Then $\Delta V^*(\sigma_s)$ is a decreasing function of 
$\sigma_s$.  

Before going any further in specifying the ingredients of 
$R_0$, $\Delta V^*$, etc., it is useful to recast the 
equations of motion in the following form.
Define
\begin{equation}
n_{tot}\equiv n_{+}+n_{-};~~~~n_{\Delta}\equiv n_{-}-n_{+};
\end{equation}
and
\[
{\cal C}(\sigma_s)  \equiv 
{1\over 2}\,\left[\exp\left(-
{\Delta V^*(\sigma_s)\over v_f}\right) + 
\,\exp\left(-
{\Delta V^*(-\sigma_s)\over v_f}\right) \right];
\]
\begin{equation}
{\cal S}(\sigma_s) \equiv
{1\over 2}\,\left[\exp\left(-
{\Delta V^*(\sigma_s)\over v_f}\right) - 
\,\exp\left(-
{\Delta V^*(-\sigma_s)\over v_f}\right) \right]. 
\end{equation}
(For convenience, and in order to be consistent with later assumptions, 
we have suppressed other possible 
arguments of the functions ${\cal C}(\sigma_s)$ and ${\cal 
S}(\sigma_s)$.)
Then (\ref{epdot}) becomes:
\begin{equation}
\label{epdot2}
\dot\varepsilon_s^{in} = 
R_0\,V_z\,\Delta\varepsilon\,\Bigl[n_{tot}\,{\cal 
S}(\sigma_s) -n_{\Delta}\, {\cal C}(\sigma_s)\Bigr].
\end{equation}
The equations of motion for $n_{\Delta}$ and $n_{tot}$ are:
\begin{equation}
\dot n_{\Delta}={2\dot\varepsilon_s^{in}\over 
V_z\,\Delta\varepsilon}\,\left(1-{\sigma_s\, n_{\Delta}\over 
\bar\sigma\,n_{\infty}}\right),
\end{equation}
and
\begin{equation}
\label{ntotdot}
\dot n_{tot}={2\,\sigma_s\,\dot\varepsilon_s^{in}\over 
V_z\,\Delta\varepsilon\,\bar\sigma}\,\left(1-{n_{tot}\over 
n_{\infty}}\right),
\end{equation}
where $\bar\sigma$ and $n_{\infty}$ are defined by
\begin{equation}
C_1\equiv {2\over 
V_z\,\Delta\varepsilon\,n_{\infty}\,\bar\sigma}; 
~~~~C_2\equiv{1\over V_z\,\Delta\varepsilon\,\bar\sigma}.
\end{equation}
From (\ref{ntotdot}), we see that $n_{\infty}$ is the stable 
equilibrium value of $n_{tot}$ so long as 
$\sigma_s\,\dot\varepsilon_s^{in}$ remains positive.  
$\bar\sigma$ is a characteristic stress that, in certain 
cases, turns out to be the plastic yield stress.  As we 
shall see, we need only the above form of the equations of 
motion to deduce the existence of the plastic yield stress 
and to compute some elementary properties of the system.

The interesting time-dependent behavior of the system, 
however, depends sensitively on the as-yet unspecified 
ingredients of these equations. Consider first the rate 
factor $R_0$.  Our zero-temperature hypothesis implies that 
$R_0$ should be zero whenever the inelastic shear rate 
$\dot\varepsilon_s^{in}$ and the elastic shear rate 
$\dot\varepsilon_s^{el} = \dot\sigma_s/2\mu$ both vanish.  
Accordingly, we assume that
\begin{equation}
\label{R0}
R_0\cong 
\nu^{1/2}\,\left[(\dot\varepsilon_s^{el})^2+(\dot\varepsilon_s^{in
})^2\right]^{1/4}, 
\end{equation} 
where $\nu$ is a constant that we must determine from the numerical 
data. Note that $\nu$ contains both an attempt frequency and a 
statistical factor associated with the multiplicity of trajectories 
leading from one state to the other in an active zone.\cite{JLanger69}

We can offer only a speculative justification for the right-hand side of 
(\ref{R0}).  The rearrangements that occur during irreversible shear 
transformations are those in which molecules deviate from the 
trajectories that they would follow if the system were a continuous 
medium undergoing affine strain. If we assume that these deviations are 
diffusive, and that the 
affine deformation over some time interval scales like the strain rate,
then the non-affine transformation rate must scale like the square root
of the affine rate.  (Diffusive deviations from smooth trajectories
have been observed directly in numerical simulations of sheared 
foams,\cite{SLanger97} but only in the equivalent of our plastic 
flow regime.)  In (\ref{R0}), we further assume that the elastic and 
inelastic strain rates are incoherent, and thus write the sum of squares 
within the brackets.  In what follows, we shall not be able to test the 
validity of (\ref{R0}) with any precision.  Most probably, the only 
properties of importance  to us for present purposes are the magnitude 
of $R_0$ and the fact that it vanishes when the shear rates vanish.  

Finally, we need to specify the ingredients of $\Delta V^*$ and $v_f$. 
For $\Delta V^*$, we choose the simple form: 
\begin{equation}
\label{V0sigma} \Delta V^*(\sigma_s)= V_0^*\,\exp(-\sigma_s/\bar\mu)
\end{equation} 
where $V_0^*$ is a volume, perhaps of order the average molecular volume 
$v_0$, and $\bar\mu$ has the dimensions of a shear modulus.  The right-
hand side
of (\ref{V0sigma}) simply reflects the fact that the free volume needed
for an activated transition will decrease if the zone in question is
loaded with a stress which coincides with the direction of the
resulting strain.  We choose the exponential rather than a linear
dependence because it makes no sense for the incremental free volume 
$V_0^*$ to be negative, even for very large values of the applied 
stresses. 

Irreversibility enters the theory via a simple
switching behavior that occurs when the $\sigma_s$-
dependence of $\Delta V^*$ in (\ref{V0sigma}) is so strong 
that it converts a negligably small rate at $\sigma_s=0$ to 
a large rate at relevant, non-zero values of $\sigma_s$.  If 
this happens, then zones that have switched in one direction 
under the influence of the stress will remain in that state 
when the stress is removed.  

In the formulation presented here, we consider $v_f$ to be constant.
This is certainly an approximation; in fact, as seen in Figure 
\ref{strain-vs-time}, the system dilates during shear deformation. We 
have experimented with  versions of this theory in which the dilation 
plays a controlling role in the dynamics via variations in $v_f$. We 
shall not discuss these versions further because they behaved in ways 
that were qualitatively different from what we observed in our 
simulations.  The differences arise from feedback between inelastic 
dilation and flow which occur in these dilational models, and apparently 
not in the simulations.  A simple comparison of the quantities involved 
demonstrates that the assumption that $v_f$ is approximately constant
is consistent with our other assumptions. If we assume that the
increment in free volume at zero stress must be of order the volume of
a small particle, $V_0^* \approx v_0 \approx 0.3$, and then look ahead
and use our best-fit value for the ratio $V_0^* / v_f \approx 14.0$ 
(see Section \ref{sec:parameters}, Table \ref{tab2}), we find
$v_f \approx 0.02$.  Since the change in free
volume due to a dilational strain $\varepsilon_d$ is $\Delta v_f =
\varepsilon_d / \rho$, where $\rho$ is the number density, and
$\varepsilon_d < 0.2\%$ for all shear stresses except those very close to
yield, it appears that, generally, 
$\Delta v_f \approx \varepsilon_d\,v_0 \ll v_f$.  Even when
$\varepsilon_d = 1\%$, the value observed in our simulations at yield,
the dilational free volume is only about the same as the initial free volume
estimated by this analysis.

As a final step in examining the underlying structure of 
these equations of motion, we make the following scaling 
transformations:
\begin{equation}
\label{scaledfirst}
{2\mu\,\varepsilon_s^{in}\over \bar{\sigma}}\equiv {\cal E}; 
~~~~{n_{\Delta}\over n_{\infty}}\equiv 
\Delta;~~~~{n_{tot}\over n_{\infty}}\equiv 
\Lambda;~~~~{\sigma_s\over \bar\sigma}\equiv \Sigma.
\end{equation}
Then we find:
\begin{equation}
\label{Edot}
\dot{\cal E}= \bar{\cal E}\,{\cal F}(\Sigma,\,\Lambda,\,\Delta);
\end{equation}
\begin{equation}
\label{Ddot}
\dot\Delta=2\,{\cal F}(\Sigma,\,\Lambda,\,\Delta)\,(1-
\Sigma\,\Delta);
\end{equation}
\begin{equation}
\label{Ldot}
\dot\Lambda=2\,{\cal 
F}(\Sigma,\,\Lambda,\,\Delta)\,\Sigma\,(1-\Lambda);
\end{equation}
where
\begin{equation}
\label{Fdef}
{\cal F}(\Sigma,\,\Lambda,\,\Delta)=R_0\,\left[\Lambda\,{\cal 
S}(\Sigma)-\Delta\,{\cal 
C}(\Sigma)\right];
\end{equation}
and:
\begin{eqnarray}
{\cal C}(\Sigma)={1\over 2}\,\left[\exp(-{V_0^* \over v_f}
e^{-A\,\Sigma}) + \exp(-{V_0^* \over v_f}
e^{A\,\Sigma})\right];\cr\nonumber\\ {\cal S}(\Sigma)={1\over
2}\,\left[\exp(-{V_0^* \over v_f} e^{-A\,\Sigma}) - \exp(-{V_0^* \over
v_f} e^{A\,\Sigma})\right]. 
\end{eqnarray}
Here,
\begin{equation} 
A\equiv{\bar\sigma \over
\bar\mu};~~~~ \bar{\cal 
E}\equiv{2\,\mu\,V_z\,\Delta\varepsilon\,n_{\infty}
\over\bar\sigma}. \label{scaledlast} 
\end{equation}
The rate factor in (\ref{R0}) can be rewritten: 
\begin{equation} 
R_0 =
\tilde\nu^{1\over2}\,\Bigl(\dot\Sigma^2+\dot{\cal E}^2\Bigr)^{1\over4},
\end{equation} 
where 
\begin{equation}
\tilde{\nu} \equiv {\bar\sigma\over 2\mu}\nu.
\end{equation}

\subsection{Special Steady-State Solutions}

Although in general we must use numerical methods to solve the fully 
time dependent equations of motion, we can solve them analytically for 
special cases in which the stress $\Sigma$ is held constant. Note that 
none of the results presented in this subsection, apart from 
(\ref{Edotsteady}), depend on our specific choice of the rate factor 
$R_0$.

There are two specially important steady-state solutions at constant 
$\Sigma$.  The first of these is a jammed solution in which $\dot{\cal 
E}=0$, that is ${\cal F}(\Sigma,\Lambda,\Delta)$ vanishes and therefore:
\begin{equation} 
\label{jamcond} 
\Delta = \Lambda
{{\cal S}(\Sigma) \over {\cal C}(\Sigma)} = \Lambda {\cal T}(\Sigma);
\end{equation} 
where
\begin{equation} 
{\cal T}(\Sigma) \equiv 1- 2\left[ 1 +
\exp\left(2 {V_0^* \over v_f} \sinh(A\,\Sigma)\right) \right]^{-1}.
\end{equation} 

Now suppose that, instead of increasing the stress at a finite rate as 
we have done in our numerical experiments, we let it jump 
discontinuously --- from zero, perhaps --- to its value $\Sigma$ at time 
$t=0$.  While $\Sigma$ is constant, (\ref{Ddot}) and (\ref{Ldot}) can be 
solved to yield:
\begin{equation} 
\label{DLrelation}
{1-\Lambda(t) \over 1 - \Lambda(0)}= {1-\Sigma \Delta(t) \over 1- \Sigma
\Delta(0)}, 
\end{equation} 
where $\Lambda(0)$ and $\Delta(0)$ denote the initial values of 
$\Lambda(t)$ and $\Delta(t)$ respectively. Similarly, we can solve 
(\ref{Edot}) and (\ref{Ddot}) for ${\cal E}(t)$ in terms of $\Delta(t)$ 
and obtain a relationship between the bias in the population of defects 
and the change in strain, 
\begin{equation} 
\label{EDrelation} 
{\cal E}(t)
= {\cal E}(0) + {\bar{\cal E} \over 2 \Sigma} \ln\left( {1 - \Sigma
\Delta(0) \over 1 - \Sigma \Delta(t)}\right). 
\end{equation} 
Combining (\ref{jamcond}), (\ref{DLrelation}) and (\ref{EDrelation}),
we can determine the change in strain prior to jamming.  That is, for 
$\Sigma$ sufficiently small that the following limit exists, we can 
compute a final inelastic strain ${\cal E}_f$:
\begin{eqnarray}
\label{Ef}
\lefteqn{{\cal E}_f \equiv \lim_{t\to \infty} {\cal E}(t) =} \nonumber \\
& & {\cal E}(0) + {\bar{\cal E} \over 2 \Sigma}\ln\left(1 +
\Sigma \,{{\cal T}(\Sigma) \Lambda(0) - \Delta(0)\over 1 - \Sigma {\cal
T}(\Sigma)}\right). 
\end{eqnarray}
The right-hand side of (\ref{Ef}), for ${\cal E}(0)=\Delta(0)=0$, should 
be at least a rough approximation for the inelastic strain as a function 
of stress as shown in Figure ~\ref{finalstrain}. 

The preceding analysis is our mathematical description of how the system 
jams due to the two-state nature of the shear transformation zones.  
Each increment of plastic deformation corresponds to the transformation 
of zones aligned favorably with the applied shear stress.  As the zones 
transform, the bias in their population --- i.e. $\Delta$ --- grows. 
Eventually, all of the favorably aligned zones that can transform at the 
given magnitude and direction of the stress have undergone their one 
allowed transformation, $\Delta$ has become large enough to cause ${\cal 
F}$ in (\ref{Fdef}) to vanish, and plastic deformation comes to a halt.

The second steady-state is a plastically flowing solution in which 
$\dot{\cal E}\ne 0$ but $\dot\Delta=\dot\Lambda=0$.  From (\ref{Ddot}) 
and (\ref{Ldot}) we see that this condition requires:
\begin{equation}
\label{flowcond}
\Delta = {1 \over \Sigma};~~~~\Lambda = 1.
\end{equation}
This leads us directly to an equation for the strain-rate at 
constant applied stress,
\begin{equation}
\label{Edotsteady}
\dot{\cal E} = \tilde{\nu}\,\bar{\cal E}^2\,\left[ {\cal S}(\Sigma) -
{1\over\Sigma}{\cal C}(\Sigma) \right]^2. 
\end{equation}
This flowing solution arises from the non-linear annihilation and 
creation terms in (\ref{ndot}).  In the flowing state, stresses are high 
enough that shear transformation zones are continuously created.  A 
balance between the rate of zone creation and the rate of transformation 
determines the rate of deformation.

Examination of (\ref{Ddot}) and (\ref{Ldot}) reveals that the jammed 
solution (\ref{jamcond}) is stable for low stresses, while the flowing 
solution (\ref{flowcond}) is stable for high stresses.  The crossover 
between the two solutions occurs when both (\ref{jamcond}) and 
(\ref{flowcond}) are satisfied.  This crossover defines the yield stress 
$\Sigma_y$, which satisfies the condition 
\begin{equation}
\label{yieldcond} 
{1 \over \Sigma_y} = {\cal T}(\Sigma_y).
\end{equation}
Note that the argument of the logarithm in (\ref{Ef}) diverges at 
$\Sigma= \Sigma_y$.  Note also that, so long as $(2V_0^* / 
v_f)\sinh(A\Sigma_y) \gg 1$, (\ref{yieldcond}) implies that 
$\Sigma_y\cong 1$.  This inequality is easily satisfied for the 
parameters discussed in the following subsection.  Thus the dimensional 
yield stress $\sigma_{y}$ is approximated accurately by $\bar\sigma$ in 
our original units defined in (\ref{scaledfirst}).  

\subsection{Parameters of the Theory}
\label{sec:parameters}

There are five adjustable system parameters in our theory: $\bar\sigma$, 
$V_z\,\Delta\varepsilon\,n_{\infty}$, $\nu$, $V_0^*/v_f$, and $\bar\mu$.  
In addition, we must specify initial conditions for ${\cal E}$, 
$\Delta$, and $\Lambda$.  For all cases of interest here, ${\cal 
E}(0)=\Delta(0)=0$.  However, $\Lambda(0)=n_{tot}(0)/n_{\infty}$ is an 
important parameter that characterizes  the as-quenched initial state of 
the system and which remains to be determined. 

To test the validity of this theory, we now must find out whether there 
exists a set of physically reasonable values of these parameters for 
which the theory accounts for all (or almost all) of the wide variety of 
time-dependent phenomena seen in the molecular-dynamics experiments.  
Our strategy has been to start with rough guesses based on our 
understanding of what these parameters mean, and then to adjust these 
values by trial and error to fit what we believe to be the crucial 
features of the experiments.  We then have used those values of the 
parameters in the equations of motion to check agreement with other 
numerical experiments. In adjusting parameters, we have looked for 
accurate agreement between theory and experiment in low-stress 
situations where we expect the concentration of active shear 
transformation zones to be low; and we have allowed larger discrepancies 
near and above the yield stress where we suspect that interactions 
between the zones may invalidate our mean-field approximation.  Our 
best-fit parameters are shown in Table~\ref{tab2}. 

\begin{minipage}{3.5in}
\begin{table}[]
\begin{center}
\begin{minipage}{2in}
 \begin{tabular}{|@{\hspace{0.2in}}c@{\hspace{0.2in}}|cc|}
 Parameter & Value &\\ \hline
 $\bar{\sigma}$ & 0.32 &\\ \hline
 $V_z\,\Delta\varepsilon\,n_\infty$ & $5.7\%$ &\\ \hline
 $\nu$ & 50.0 &\\ \hline
 $V^*_0 / v_f$ & 14.0 &\\ \hline
 $\bar{\mu}$ & 0.25 &\\ \hline
 $n_{tot}(0) / n_\infty$ & 2.0 &\\
 \end{tabular}
\end{minipage}
\end{center}
\begin{minipage}{8.1cm}
 \caption{\label{tab2}
 Values of parameters for comparison to simulation data}
\end{minipage}
\end{table}
\end{minipage}

The easiest parameter to fit should be $\bar\sigma$ because it should be 
very nearly equal to the yield stress.  That is, it should be somewhere 
in the range 0.30-0.35 according to the data shown in 
Fig.~\ref{finalstrain}.  Note that we cannot use (\ref{Ef}) to fit the 
experimental data near the yield point because both the numerical 
simulations and the theory tell us that the system approaches its 
stationary state infinitely slowly there.  Moreover, we expect 
interaction effects to be important here.  The solid curve in 
Fig.~\ref{finalstrain} is the theoretically predicted strain found by 
integrating the equations of motion for 800 time units, the duration of 
the longest of the simulation runs.  The downward adjustment of 
$\bar\sigma$, from its apparent value of about 0.35 to its best-fit 
value of 0.32, has been made on the basis of the latter time-dependent 
calculations plus evidence about the effect of this parameter in other 
parts of the theory.  

Next we consider $V_z\,\Delta\varepsilon\,n_\infty$, a dimensionless 
parameter which corresponds to the amount of strain that would occur if 
the density of zones were equal to the equilibrium concentration 
($n_{tot} = n_\infty$) and if all the zones transformed in the same 
direction in unison.  Alternatively, if the local strain increment 
$\Delta\varepsilon$ is about unity, then this parameter is the fraction 
of the volume of the system that is occupied by shear transformation 
zones.  In either way of looking at this quantity, our best-fit value of  
5.7\% seems sensible.

The parameter $\nu$ is a rate which is roughly the product of an attempt 
frequency and a statistical factor.  The only system-dependent quantity 
with the dimensions of inverse time is the molecular vibrational 
frequency, which we have seen is of order unity.  Our best-fit value of 
50 seems to imply that the statistical factor is moderately large which, 
in turn, implies that the shear transformation zones are fairly complex, 
multi-molecule structures.  Lacking any first-principles theory of this 
rate factor, however, we cannot be confident about this observation.

Our first rough guess for a value of $V_0^*/v_f$ comes from the 
assumption that $\Delta V^*$ must be about one molecular volume in the 
absence of an external stress, and that $v_f$ is likely to be about a 
tenth of this. Thus our best-fit value of 14.0 is reassuringly close to 
what we expected. 

The parameter $\bar\mu$, a modulus that characterizes the sensitivity of 
$\Delta V^*$ to the applied stress, is especially interesting.  Our 
best-fit value of 0.25 is almost two orders of magnitude smaller than a 
typical shear modulus for these systems.  This means that the shear 
transformations are induced by relatively small stresses or, 
equivalently, the internal elastic modes within the zones are very soft.  
This conclusion is supported quite robustly by our fitting procedure. 
Alternative assumptions, such as control by variations in the average 
free volume $v_f$ discussed earlier, produce qualitatively wrong 
pictures of the time dependent onset of plastic deformation.

Finally, we consider $\Lambda(0)= n_{tot}(0)/n_\infty$, the ratio of the 
inital zone density to the equilibrium zone density. This parameter 
characterizes the transient behavior associated with the initial quench; 
that is, it determines the as-quenched system's first response to an 
applied stress.  We can learn something about this parameter by looking 
at later behavior, i.e. the next few segments of a hysteresis loop like 
that shown in Figure~\ref{cycle}.  If, as is observed there, the loop 
narrows after the first leg, then we know that there was an excess of 
shear transformation zones in the as-quenched system, and that this 
excess was reduced in the initial deformation.  An initial excess means 
$\Lambda(0)>1$, consistent with our best-fit value of 2.0.

\subsection{Comparisons between Theory and Simulations}

We now illustrate the degree to which this theory can --- and cannot --- 
account for the phenomena observed in the numerical experiments.  

\begin{figure}
\epsfxsize=3.0in
\centerline{\epsfbox{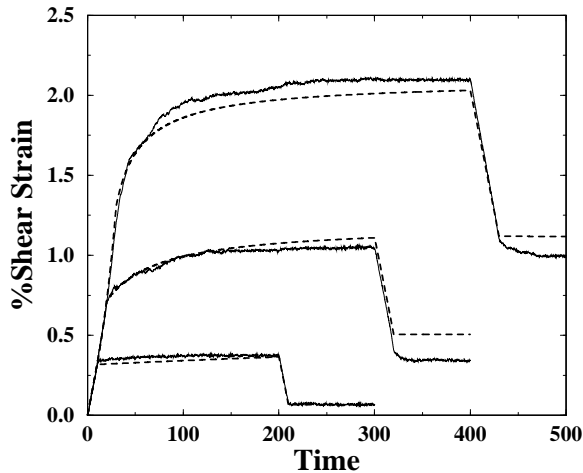}}
\begin{minipage}[t]{8.1cm} 
\caption{Strain vs. time for simulations in which the
stress has been ramped up at a controlled rate to stresses of 0.1, 0.2 and
0.3, held constant, and then ramped down to zero (solid lines).  The 
dashed lines are the corresponding theoretical 
predictions.}\label{modelstr} 
\end{minipage} 
\end{figure}

Figure \ref{modelstr} summarizes one of the principal successes of the 
theory, specifically, its ability to predict the time-dependent onset of 
plastic deformation over a range of applied stresses below the yield 
stress. The solid lines in the Figure show the shear strains in three 
different simulations as functions of time. In each simulation the 
stress is ramped up at the same controlled rate, held constant for a 
period of time, and then ramped down, again at the same rate.  In the 
lowest curve the stress reaches a maximum of 0.1 in
our dimensionless stress units ($e_{SL}/a_{SL}^2$), in the middle curve
0.2, and in the highest 0.3.  The dashed lines show the predictions of 
the theory.  The excellent agreement during and after the ramp-up is our 
most direct evidence for the small value of $\bar\mu$ quoted above.  The 
detailed shapes of these curves at the tops of the ramps, where 
$\dot\sigma_s$ drops abruptly to zero, provide some qualitative support 
for our choice of the rate dependence of $R_0$ in (\ref{R0}).  As shown 
in Figure \ref{finalstrain} and discussed in the preceding subsection, 
the final inelastic strains in these ramp-up experiments are also 
predicted adequately by the theory. 

The situation is different for the unloading phases of these 
experiments, that is, during and after the periods when the stresses are 
ramped back down to zero.  The theoretical strain rates shown in Figure 
\ref{modelstr} vanish abruptly at the bottoms of the ramps because our 
transformation rates become negligably small at zero stress.  In the two 
experimental curves for the higher stresses, however, the strain 
continues to decrease for a short while after the  
\noindent stresses have stopped 
changing.  Our theory seems to rule out any such recovery of inelastic 
strain at zero stress; thus we cannot account for this phenomenon except 
to remark that it must have something to do with the initial state of 
the as-quenched system.  As seen in Figure \ref{cycle}, no such recovery 
occurs when the system is loaded and unloaded a second time.  

In Figure \ref{cycle}, we compare the stress-strain hysteresis loop in 
the simulation (solid line) with that predicted by the theory (dashed 
line). Apart from the inelastic strain recovery after the first 
unloading in the simulation, the theory and the experiment agree well 
with one another at least through the reverse loading to point {\bf k}.  
The agreement becomes less good in subsequent cycles of the hysteresis 
loop, possibly because shear bands are forming during repeated plastic 
deformations.

In the last of the tests of theory to be reported here, we have added in 
Figure \ref{strainrate} two theoretical curves for stresses as 
functions of strain at the two different constant strain rates used in 
the simulations.  The agreement between theory and experiment is better 
than we probably should expect for situations in which the stresses 
necessarily rise to values at or above the yield stress.  Moreover, the 
validity of the comparison is obscured by the large fluctuations in the 
data, which we believe to be due primarily to small sample size.  

Among the interesting features of the theoretical results in Figure 
\ref{strainrate} are the peaks in the stresses that occur just prior to 
the establishment of steady-states at constant stresses.  These peaks 
occur because the internal degrees of freedom of the system, 
specifically $\Delta(t)$ and $\Lambda(t)$, cannot initially equilibrate 
fast enough to accomodate the rapidly increasing inelastic strain.  Thus 
there is a transient stiffening of the material and a momentary increase 
in the stress needed to maintain the constant strain rate.  This kind of 
effect may in part be the explanation for some of the oscillations in 
the stress seen in the experiments.  In a more speculative vein, we note 
that this is our first direct hint of the kind of dynamic plastic 
stiffening that is needed in order to transmit high stresses to crack 
tips in brittle fracture.  The orders of magnitude of the time scales 
are even roughly the same.  The strain rates used here, and 
those that occur near the tips of brittle cracks, are both of the order 
of $10^7$ per second.

\section{Concluding Remarks}

The most striking and robust conclusion to emerge from this 
investigation, in our opinion, is that a wide range of realistic, 
irreversible, viscoplastic phenomena occur in an extremely simple 
molecular-dynamics model --- a two-dimensional, two-component, Lennard-Jones 
amorphous solid at essentially zero temperature.  An almost equally
striking conclusion is that a theory based on 
the dynamics of two-state shear transformation zones is in substantial 
agreement with the observed behavior of this model.  This theory has 
survived several quantitative tests of its applicability.

We stated in our Introduction that this is a preliminary report.  Both 
the numerical simulations and the theoretical analysis require careful 
evaluation and improvements.  Most importantly, the work so far raises 
many important questions that need to be addressed in future 
investigations.  

The first kind of question pertains to our molecular-dynamics 
simulations: Are they accurate and repeatable?  We believe that they are 
good enough for present purposes, but we recognize that there are 
potentially important difficulties.  The most obvious of these is that 
our simulations have been performed with very small systems; thus, size 
effects may be important.  For example, the fact that only a few shear 
transforming regions are active at any time may account for abrupt jumps 
and other irregularities sometimes seen in the simulations, e.g. in 
Figure \ref{strainrate}. We have performed the simulations in a periodic 
cell to eliminate edge effects.  We also have tried to compare results 
from two systems of different sizes, although only the results from the larger
system are presented here.  Unfortunately, comparisons between 
any two different initial configurations are difficult because of our 
inability, as yet, to create reproduceable glassy starting 
configurations (a problem which we shall discuss next).  However, we 
have seen qualitatively the same  behavior in both systems, and assume 
that phenomena which are common to both systems can be used as a guide 
for theoretical investigations.  

As noted in Section IIB and in Table \ref{tab1}, our two systems had 
quite different elastic moduli.  (Remarkably, their yield stresses were 
nearly identical. It would be interesting to learn whether this is a 
repeatable and/or physically important phenomenon.)  The discrepancy 
between the elastic properties of the two systems leads us to believe 
that, in future work, we shall have to learn how to control the initial 
configurations more carefully, perhaps by annealing the systems after 
the initial quenches.  Unfortunately, straightforward annealing at 
temperatures well below the glass transition is not yet possible with 
standard molecular-dynamics algorithms, which can simulate times only up 
to about a microsecond for systems of this size even with today's 
fastest computers.  Monte Carlo techniques or accelerated molecular-dynamics 
algorithms may eventually be useful in this effort.\cite{Barkema96,Voter97a,Voter97b}  
An alternative strategy may be simply to look at larger numbers of 
simulations.

By far the most difficult and interesting questions, however, pertain to 
our theoretical analysis.  Although Figures \ref{STzones1} and 
\ref{STzones2} provide strong evidence that irreversible shear 
transformations are localized events, we have no sharp 
definition of a ``shear transformation zone.''  So far, we have 
identified these zones only after the fact, that is, only by observing 
where the transformations are taking place.  Is it possible, at least in 
principle, to identify zones before they become active?  

One ingredient of a better definition of shear transformation zones will 
be a generalization to isotropic amorphous systems in both two and three 
dimensions.  As we noted in Chapter III, our functions $n_{\pm}(t)$ 
should be tensor quantities that describe distributions over the ways in 
which the individual zones are aligned with respect to the orientation 
of the applied shear stress. We believe that this is a relatively easy 
generalization; one of us (MLF) expects to report on work along these 
lines in a later publication.  

Our more urgent reason for needing a better understanding of shear 
transformation zones is that, without such an understanding, we shall 
not be able to find first-principles derivations of several, as-yet 
purely phenomenological, ingredients of our theory.  It might be useful, 
for example, to be able to start from the molecular force constants and 
calculate  the parameters $V_0^*$ and $\bar\mu$ that occur in the 
activation factor (\ref{V0sigma}).  These parameters, however, seem to 
have clear physical interpretations; thus we might be satisfied to 
deduce them from experiment.  In contrast, the conceptually most 
challenging and important terms are the  rate factor in (\ref{R0}) and 
the annihilation and creation terms in (\ref{ndot}), where we do not 
even know what the functional forms ought to be.

Calculating the rate factor in (\ref{R0}), or a correct version of that 
equation, is clearly a very fundamental problem in nonequilibrium 
statistical physics.  So far as we know, there are no studies in the 
literature that might help us compute the force fluctuations induced at 
some site by externally driven deformations of an amorphous material.  
Nor do we know how to compute a statistical prefactor analogous, 
perhaps, to the entropic factor that converts an activation energy to an 
activation free energy.\cite{JLanger69}  We do know, however, that that 
entropic factor will depend strongly on the size and structure of the 
zone that is undergoing the transformation.

As emphasized in Chapter III, the annihilation and creation terms in 
(\ref{ndot}) describe interaction effects.  Even within the framework of 
our mean-field approximation, we do not know with any certainty what 
these terms should be.  Our assumption that they are proportional to the 
rate of irreversible work is by no means unique.  (Indeed, we have tried 
other possibilities in related investigations and have arrived at 
qualitatively similar conclusions.) Without knowing more about the 
nature of the shear transformation zones, it will be difficult to derive 
such interaction terms from first principles. 

A better understanding of these interaction terms is especially 
important because these are the terms that will have to be modified when 
we go beyond the mean-field theory to account for correlations between 
regions undergoing plastic deformations.  We know from our simulations 
that the active zones cluster even at stresses far below the plastic 
yield stress; and we know that plastic yield in real amorphous materials 
is dominated by shear banding. Thus, generalizing the present mean-field 
theory to one which takes into account spatial variations in the 
densities of shear transformation zones must be a high priority in this 
research program.

Finally, we return briefly to the questions which motivated this 
investigation: How might the dynamical effects described here, which 
must occur in the vicinity of a crack tip, control crack stability and 
brittle/ductile behavior?  As we have seen, our theoretical picture of 
viscoplasticity does allow large stresses to be transmitted, at least 
for short times, through plastically deforming materials.  It should be 
interesting to see what happens if we incorporate this picture into 
theories of dynamic fracture. 

\section*{Acknowledgements}

This research was supported by DOE Grant No. DE-FG03-84ER45108, by the 
DOE Computational Sciences Graduate Fellowship Program and, in part, 
by the MRSEC Program of the National Science Foundation under 
Award No. DMR96-32716.

We wish particularly to thank Alexander Lobkovsky for his attention to 
this project and for numerous useful suggestions.  We also thank A. 
Argon and F. Lange for guidance in the early stages of this work, J. 
Cahn for directing us to the papers of E. Hart, and S. Langer and A. Liu 
for showing us their closely related results on the dynamics of sheared 
foams. 
\section*{ }
\bibliography{biblio/jam}
\end{multicols}
\end{document}